\documentclass[superscriptaddress,showpacs,aps,twocolumn,amsmath,amssymb,rsi]{revtex4-1}

\usepackage{graphicx}
\usepackage{dcolumn}
\usepackage{bm}
\usepackage{textcomp}
\usepackage{verbatim} 

\usepackage{datetime}

\usepackage{color}
\definecolor{darkblue}{rgb}{0,0,0.6}
\usepackage[linkcolor=darkblue, citecolor=darkblue,colorlinks=true, breaklinks=true]{hyperref}

\begin{document}

\title{Time-domain Brillouin scattering for the determination of laser-induced \\ temperature gradients in liquids}

\author{Ievgeniia Chaban}
\email{ievgeniia.chaban@univ-lemans.fr}
\affiliation{Institut Mol\'ecules et Mat\'eriaux du Mans, UMR CNRS 6283, 
Universit\'e du Maine, 72085 Le Mans, France}
\author{Hyun D. Shin}
\affiliation{Department of Chemistry, Massachusetts Institute of Technology, Cambridge, MA 02139, USA}
\author{Christoph Klieber}
\altaffiliation[Current address: ]{EP Schlumberger, 1 rue Henri Becquerel, 92140 Clamart, France}
\affiliation{Department of Chemistry, Massachusetts Institute of Technology, Cambridge, MA 02139, USA}
\author{R\'emi Busselez}
\affiliation{Institut Mol\'ecules et Mat\'eriaux du Mans, UMR CNRS 6283, 
Universit\'e du Maine, 72085 Le Mans, France}
\author{Vitalyi E. Gusev}
\affiliation{Laboratoire d'Acoustique de l'Universit\'e du Maine, UMR CNRS 6613, 
Universit\'e du Maine, 72085 Le Mans, France}
\author{Keith A. Nelson}
\email{kanelson@mit.edu}
\affiliation{Department of Chemistry, Massachusetts Institute of Technology, Cambridge, MA 02139, USA}
\author{Thomas Pezeril}
\email{thomas.pezeril@univ-lemans.fr}
\affiliation{Institut Mol\'ecules et Mat\'eriaux du Mans, UMR CNRS 6283, 
Universit\'e du Maine, 72085 Le Mans, France}

\date{\today}

\begin{abstract}

We present an optical technique based on ultrafast photoacoustics to precisely determine the local temperature distribution profile in liquid samples in contact with a laser heated optical transducer. This ultrafast pump-probe experiment uses time-domain Brillouin scattering (TDBS) to locally determine the light scattering frequency shift. As the temperature influences the Brillouin scattering frequency, the TDBS signal probes the local laser-induced temperature distribution in the liquid. We demonstrate the relevance and the sensitivity of this technique for the measurement of the absolute laser-induced temperature gradient of a glass forming liquid prototype, glycerol, at different laser pump powers - i.e. different steady state background temperatures. Complementarily, our experiments illustrate how this TDBS technique can be applied to measure thermal diffusion in complex multilayer systems in contact to a surrounding liquid.

\end{abstract}

\maketitle

\section{INTRODUCTION}

Thermal transport has been investigated in a variety of materials in the frame of ultrafast science, ranging from metals through the investigation of electron-phonon coupling, semi-conductors, insulators, gases, liquids and at solid-liquid-gas interfaces. Such investigations are often based on a femtosecond pump-probe technique in which the change in reflectivity of the sample following the partial absorption of a femtosecond laser pulse is monitored by a time delayed probe which is optically sensitive to the transient change in sample temperature. Recorded waveforms are then analyzed and compared to models to retrieve the thermal characteristics of the sample, such as thermal conduction, diffusivity or even Kapitza interfacial thermal resistance \cite{Cahill2004, Schmidt2008, Schmidt2008b, Cahill2014, Schmidt2009}. Similarly, time resolved photoacoustics implies ultrafast lasers to optically excite and detect ultrasound in a variety of materials. In this situation, well suited for the investigation of mechanical properties of ultrathin solid or liquid samples, the occurrence of time-domain Brillouin scattering (TDBS) phenomena in transparent materials has been used to investigate viscoelastic properties of matter at ultrasound GHz to THz frequencies \cite{Lin1991}. Recently, several examples have shown that TDBS is sensitive to diverse phenomena, including spatial mechanical, optical or acousto-optical inhomogeneities \cite{Mechri2009, Steigerwald, Lomonosov2012, Nikitin2015, Dehoux2015, Danworaphong, Perez}, non-linear acoustic waves or weak shock waves \cite{Klieber2015, Bojahr2012}, and even GHz transverse acoustic phonons in viscoelastic liquids \cite{pezeril09, pezeril12, KHP+13, pezeril16}.

In this paper, since temperature influences viscoelasticity, we demonstrate through TDBS measurements the intricate coupling of longitudinal acoustic phonons to thermal properties of the investigated liquids. We present a tabletop pump-probe method which enables the measurement through TDBS of the absolute laser-induced temperature gradient in liquids in contact to an optical transducer. In the following, we will illustrate the performances of TDBS as a specific contactless local viscoelastic and temperature sensor in case of glycerol, which is a well known and well characterized prototypical glass forming liquid. Similar results were obtained for octamethylcyclotetrasiloxane (OMCTS), chosen as a confined liquid prototype, with the interest for future experiments related to molecular confinement in ultrathin liquid layers where thermal effects remains elusive.

\textcolor{black}{\section{Technique}}               

\subsection{Experimental methods}     

\begin{figure}[t!]
\centering
\includegraphics[width=9cm]{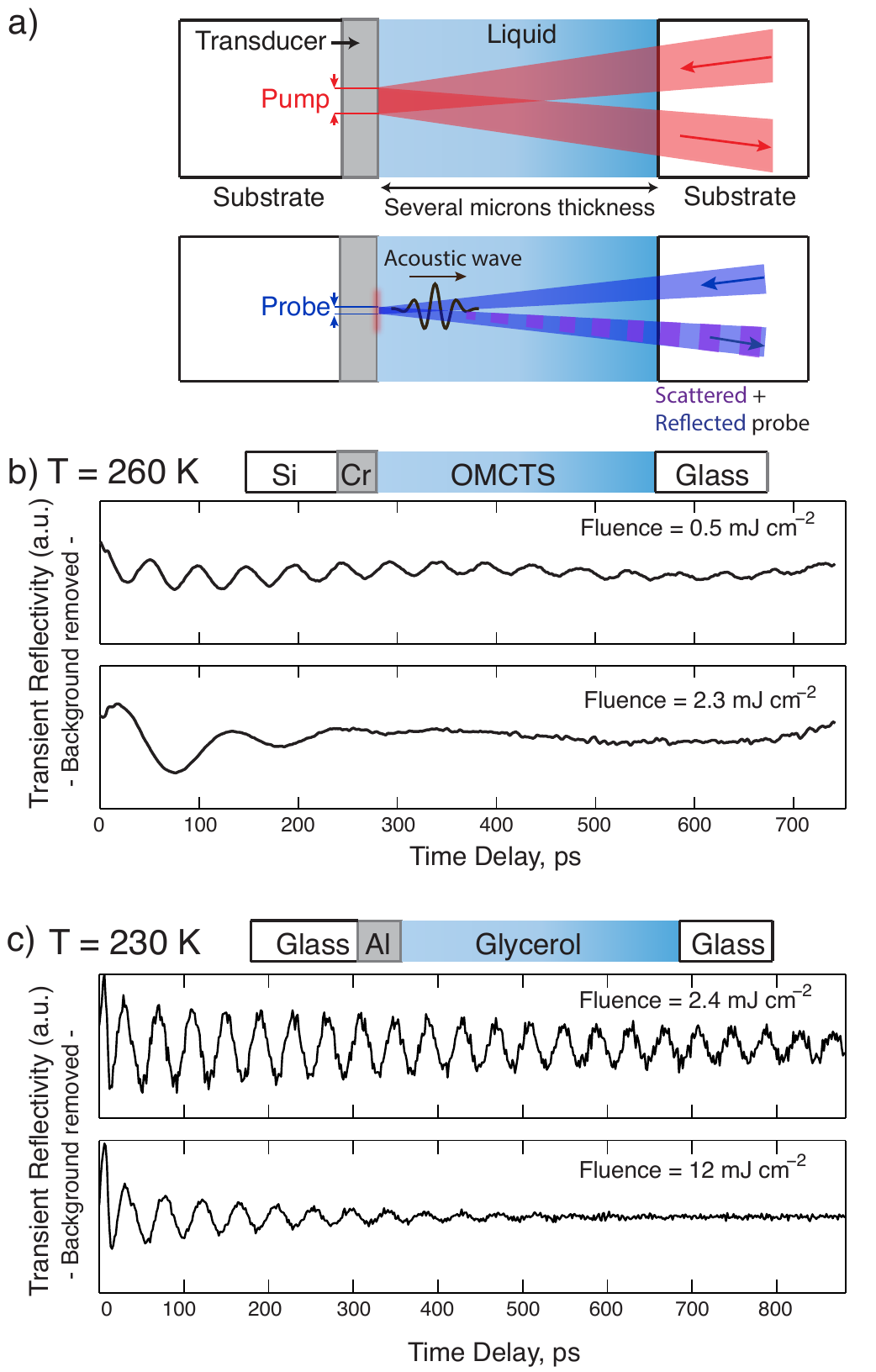}
\caption[Fig1]{(Color online) (a) The liquid is squeezed between two flat substrates, one of them being coated with a metallic transducer thin film. Upon laser irradiation by the excitation pump pulse, the acoustic waves are launched at the metallic transducer film and transmitted through the transparent liquid where they are detected through TDBS by a delayed laser probe pulse. Recorded transient reflectivity data recorded for OMCTS (b) and glycerol (c) at several different pump laser fluences - i.e cumulative thermal heating temperatures - and at different temperatures of the cryostat. The Brillouin scattering frequency drastically changes in case of OMCTS from a fivefold change in laser fluence, which indicates a solid to liquid transition mediated by the laser thermal heating. In case of glycerol, the Brillouin frequency changes by over 10\% for a fivefold increase in laser pump fluence.}
\label{fig1}
\end{figure}  

Samples were prepared by squeezing the liquid under study between two flat substrates, one of them holding a metallic thin film in contact with the liquid, as sketched in Fig. \ref{fig1}(a). The metallic thin film serves as a photoacoustic transducer, it is composed, depending on the experimental configuration, either of a 40~nm chromium film or a 80~nm aluminium film. The liquid thickness was well over 10~microns. The studied liquids were glycerol (Acros Organics$^\text{\textregistered}$, 99+\% purity) and OMCTS (Fluka$^\text{\textregistered}$, 99+\% purity), forced through several linked 0.2~$\mu$m teflon millipore filters to remove dust particles prior mounting into the liquid sample holder cell. The liquid sample cell was then transferred to a cryostat and the sample chamber was immediately evacuated. At each temperature, the sample was given sufficient time to equilibrate before data acquisition. A Peltier temperature sensor was attached on the sample holder mount, about 1~to~2~cm away from the experimental volume.

The optical experiment is based on the TDBS technique, suitable for the investigation of the frequency and temperature dependences of ultrafast acoustic dynamics in liquids at GHz frequencies \cite{pezeril09, Kli10, pezeril12, pezeril16, KHP+13, Shelton2005}. Measurements were made using an ultrafast optical pump-probe experimental setup as illustrated in Fig. \ref{fig1}(a). The laser pulses originates from a femtosecond Ti-Sapphire Coherent RegA 9000 regenerative amplifier operating at central wavelength of 790~nm and delivering 160~fs pulses at a repetition rate of 250~kHz. The laser pulses were split into two beams with the 790~nm pump beam synchronously modulated at 50 kHz frequency by an acousto-optic modulator (AOM) and focused on the surface of a metallic photo-acoustic transducer film with a gaussian spatial beam profile of FWHM $\sim$100~$\mu$m. The second, a much less energetic beam, the probe, was frequency doubled to 395 nm by a BBO crystal, time delayed, tightly focused at normal incidence on the sample surface with a spot size smaller than 20 $\mu$m, and spatially overlapped with the pump spot. The reflected probe beam was directed to a photodiode coupled to a lock-in amplifier synchronized to the 50~kHz pump modulation frequency to measure transient differential reflectivity $\Delta$R(t) as a function of time delay between pump and probe beams. Upon transient absorption of the 790~nm pump pulse over the optical skin depth of the metallic thin film, laser excited acoustic pulses were transmitted across the interface into the adjacent transparent liquid. The out-of-plane acoustic propagation of the strain pulses in the transparent liquid medium leads to the occurrence of TDBS oscillations in the transient reflectivity signal, see Fig. \ref{fig1}(b) and (c). As in any Brillouin scattering process, the frequency $\nu$ of these oscillations is related to the ultrasound velocity $v$ of the liquid, to the probe wavelength $\lambda$, to the refractive index $n$ of the medium, and to the back-scattering angle $\theta$ through
\begin{equation}\label{Brillouin}
\nu = 2 \ n \ v \cos \theta / \lambda.
\end{equation}
Many external conditions, such as the ambient temperature, influence the acoustic velocity and the index of refraction of the liquid and, as a consequence, the TDBS itself. Therefore TDBS is sensitive to a local temperature modification of the scattering liquid medium. In fact, the Brillouin scattering frequency detected in the time domain reflectivity signal monitors any change of the temperature distribution, which appears as a modification of the detected Brillouin oscillation frequency. Typically, given a speed of sound in the range of $1000-3000$~m/s in the studied liquid and a Brillouin frequency in the range of $10-30$~GHz, the characteristic TDBS sensitivity length is in the range of a couple of hundreds of nanometers. It means that any change of the overall temperature or temperature distribution in a liquid volume as small as tens of pico-liters (1000~nm$\times$probe spot surface of 100~$\mu$m FWHM diameter) can be detected from TDBS.

As in any optical pump-probe experiment, the laser pump pulse can cause permanent or irreversible sample modification at a given fluence threshold. This effect can be experimentally observed once the recorded data become fluence dependent such as the excitation of shock waves at high laser fluences \cite{Klieber2015} which reveals the departure from the linear acoustic regime to the non-linear acoustic regime. It can be a consequence as well of a local temperature rise caused by cumulative heating of the sample from the multiple laser pump pulses which brings the sample into a steady state temperature regime correlated to the laser pump fluence \cite{Cahill2004, Schmidt2008, Schmidt2008b, Cahill2014, Schmidt2009}. \textcolor{black}{Fig. \ref{fig1}(b) and (c)} displays recorded data obtained in OMCTS and glycerol at a temperature of 260~K and 230~K, respectively, as indicated by the Peltier temperature sensor, at different laser pump fluences. As seen in Fig. \ref{fig1}(b), a fivefold change in the laser pump fluence induces a drastic change in the Brillouin oscillations frequency in OMCTS, from 20.5~GHz to 8.8~GHz, and in the attenuation rate. \textcolor{black}{Since the general trend of any material is such that it is stiffer in the solid state than in the liquid state, we assume that the 20.5~GHz high frequency Brillouin scattering corresponds to the OMCTS crystalline phase and that the lower 8.8~GHz frequency corresponds to the liquid phase, see Fig. \ref{fig1}(b). This is consistent with the observation that the cumulative heating of the multiple laser pump pulses could bring OMCTS from its solid state to its liquid state from a fivefold increase in the laser pump fluence. In case of glycerol, the Brillouin frequency changes slightly from 25.1~GHz to 22.8~GHz, for a fivefold modification of the laser pump fluence. It indicates a monotonous temperature modification of glycerol which remains in its supercooled liquid phase.} The corresponding attenuation rate evolves with a modification of the laser fluence as well, however, its pertinence is out of scope of the manuscript which focus mainly on the analysis of the Brillouin frequency versus laser fluence or temperature. 

It is important to keep in mind that for the appearance of shock waves as in \cite{Klieber2015}, the Brillouin frequency tends to increase with an increase of the laser pump fluence, which is opposite to our current experimental observations. Therefore, we have neglected the effect of non-linear shock waves and solely assumed cumulative laser heating as the main mechanism responsible for the evolution of the Brillouin frequency in our present measurements. In the following, we will describe how to calibrate the measured Brillouin frequency in the studied liquid in order to employ TDBS as a specific temperature sensor.

\subsection{Temperature calibration}      

\begin{figure}[t!]
\centering
\includegraphics[width=9cm]{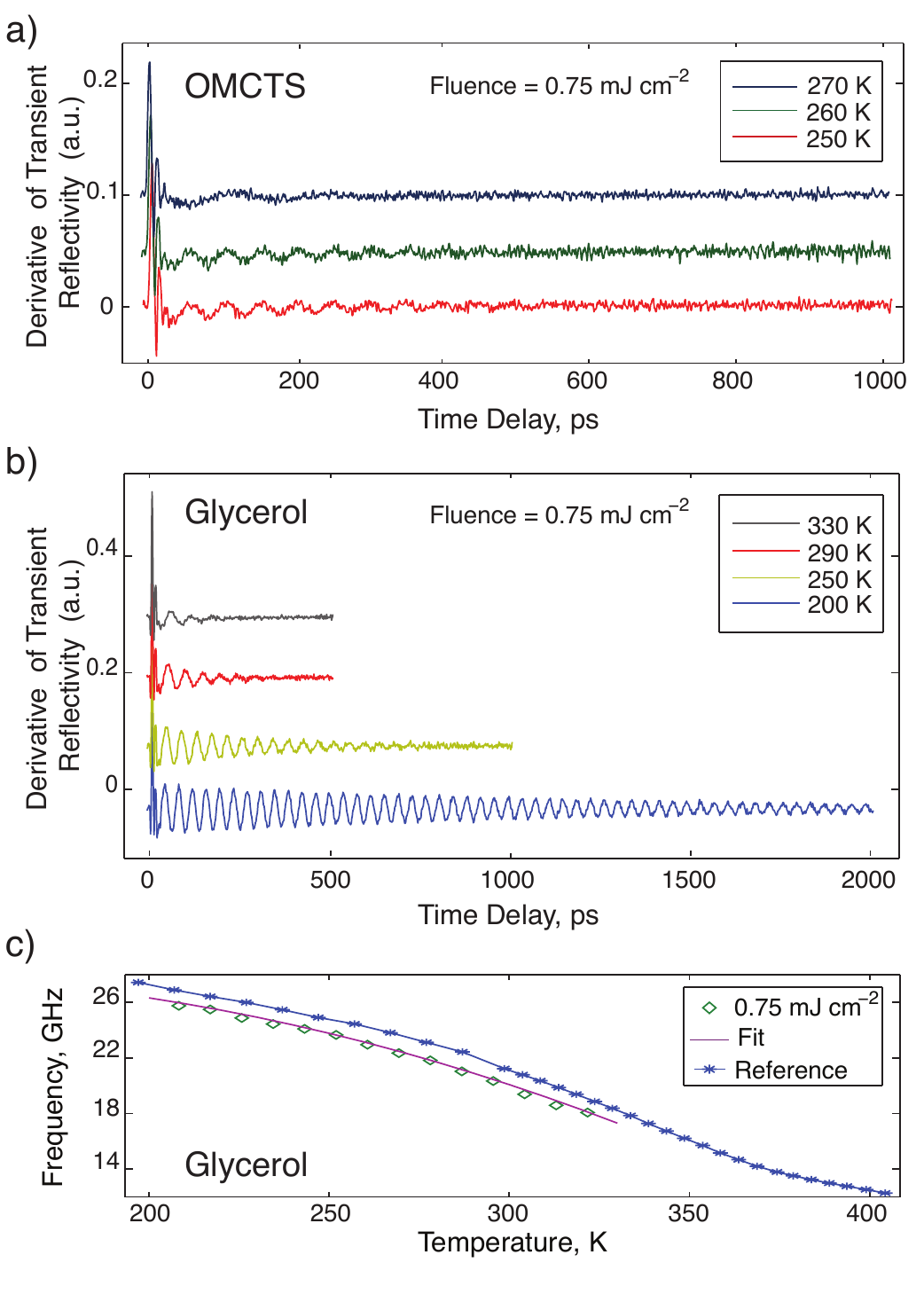}
\caption[Fig2]{(Color online) Time derivative of recorded transient reflectivity signals obtained in (a) OMCTS and in (b) glycerol at different temperatures of the cryostat and at a given laser pump fluence of 0.75 mJ.cm$^{-2}$. At such relatively low fluence, the overheating caused by the multiple laser pump pulses is moderate. (c) Temperature dependent Brillouin frequency in glycerol, from our measurements, and calculated from Comez et al. \cite{Comez2003} and Klieber et al. \cite{KHP+13}. The temperature calibration curves displayed in (c) can be used to estimate the absolute liquid temperature from the measured Brillouin frequency.}
\label{fig2}
\end{figure}

The temperature calibration measurement takes advantage of the fact that the Brillouin frequency in a liquid is strongly temperature dependent. Therefore, the Brillouin frequency can be used as a very sensitive probe to determine the absolute temperature in the experimentally investigated local region of the liquid. Since the laser pump itself can affect the Brillouin frequency at high fluences, the measurement data shown in figure~\ref{fig2} were obtained at sufficiently low pump fluence such that the effect of cumulative heating is moderate. Figure~\ref{fig2} shows the derivative of the recorded transient reflectivity change recorded at several temperatures in OMCTS and glycerol, with a sample structure as sketched in Fig.~\ref{fig1}(a), composed of a silicon substrate holding a 40~nm chromium film in contact with the liquid and a transparent cover glass substrate. The sample temperature was measured by a Peltier temperature sensor attached to the sample mount. In both cases, for both liquids, the temperature influences the Brillouin frequency and the attenuation rate, as shown in Fig.~\ref{fig2}(a) and (b). \textcolor{black}{As can be seen in Fig.~\ref{fig2}(a), OMCTS exhibits an abrupt Brillouin frequency variation at the vicinity of 260~K, which can be assigned to a sharp crystalline to liquid phase transition \cite{Hoffman, Niepmann, Sterczynska}. In case of glycerol which is an intermediate fragile glass former with a smooth glass transition temperature around T$_g$~=~186~K, the Brillouin frequency variation with temperature in Fig.~\ref{fig2}(b) is monotonous.} For this technical reason and owing to the fact glycerol is a well characterized material with readily available temperature dependent physical parameters in the literature, we have performed a meticulous temperature calibration for glycerol only. For each temperature measurement in glycerol such as the ones displayed in Fig.~\ref{fig2}(a), the frequency $\nu$ of the Brillouin oscillations observed in the transient reflectivity signal were fitted following a sinusoidal damped function in the form
\begin{equation}\label{Brillouin}
\Delta\text{R}(t) \sim \sin (2\pi\nu t + \phi) \ \exp(-\Gamma t),
\end{equation}
$\phi$ being a phase parameter. As indicated in Fig. \ref{fig2}(c), the relevant Brillouin scattering frequency fit parameter $\nu$ changes significantly as a function of temperature. The experimental temperature calibration of the Brillouin frequency has been further fitted by a smooth polynomial function in order to extract an even more reliable temperature behavior of the Brillouin scattering frequency of glycerol under our experimental conditions at 395~nm probe wavelength and a normal incidence scattering angle. Since we did not have the possibility to record the Brillouin scattering frequency at higher temperatures than 320~K, we have used a calculated calibration curve from literature data obtained from Comez et al. \cite{Comez2003} and Klieber et al. \cite{KHP+13} as a reference function to link the Brillouin frequency to the absolute glycerol temperature. \textcolor{black}{Note that the twofold change in Brillouin frequency with temperature in Fig.~\ref{fig2}(c) is mainly due to the significant change in speed of sound over the examined temperature range (200~K - 400~K) \cite{Comez2003}, rather than the change of the index of refraction which varies only a few percent over the same temperature range \cite{KHP+13}.} A slight Brillouin frequency shift of about 0.8~GHz in between our experimental data and the calculated calibration curve can be observed in Fig.~\ref{fig2}(c), which we attribute to a slight temperature shift of about 10~K caused by laser cumulative heating of the liquid, even at this relatively low pump fluence, or to a slight variation of the water content in the different glycerol samples. This discrepancy does not affect, in any way, the interpretation of the steady state cumulative thermal heating effects detailed in the following.

\textcolor{black}{The temperature sensitivity of the TDBS measurements can be estimated and expressed in terms of the noise equivalent temperature (NET) coefficient which indicates the temperature sensitivity or accuracy for a given acquisition time or frequency bandwidth of the TDBS measurements. We estimate the NET coefficient in the range of 1.5~K/$\sqrt{\text{Hz}}$. Since a single scan at a given fixed temperature of the cryostat takes about 5 seconds which corresponds, according to the Nyquist-Shannon sampling theorem to a frequency bandwidth of 0.1 Hz, the corresponding temperature accuracy is of 1.5~K $\times \sqrt{0.1} \sim$ 0.5~K. In other words, TDBS measurements can monitor dynamic temperature changes of 0.5~K at a sampling frequency of 0.1~Hz. In order to further improve the temperature accuracy of our TDBS measurements, we routinely averaged about 20 scans and the resulting temperature accuracy is within 0.1~K (1.5 K $\times \sqrt{2} \sim$ 0.1 K).}

\section{Cumulative thermal heating}               

\subsection{Different sample configurations at different laser pump fluences}   

\begin{figure}[t!!]
\centering
\includegraphics[width=9cm]{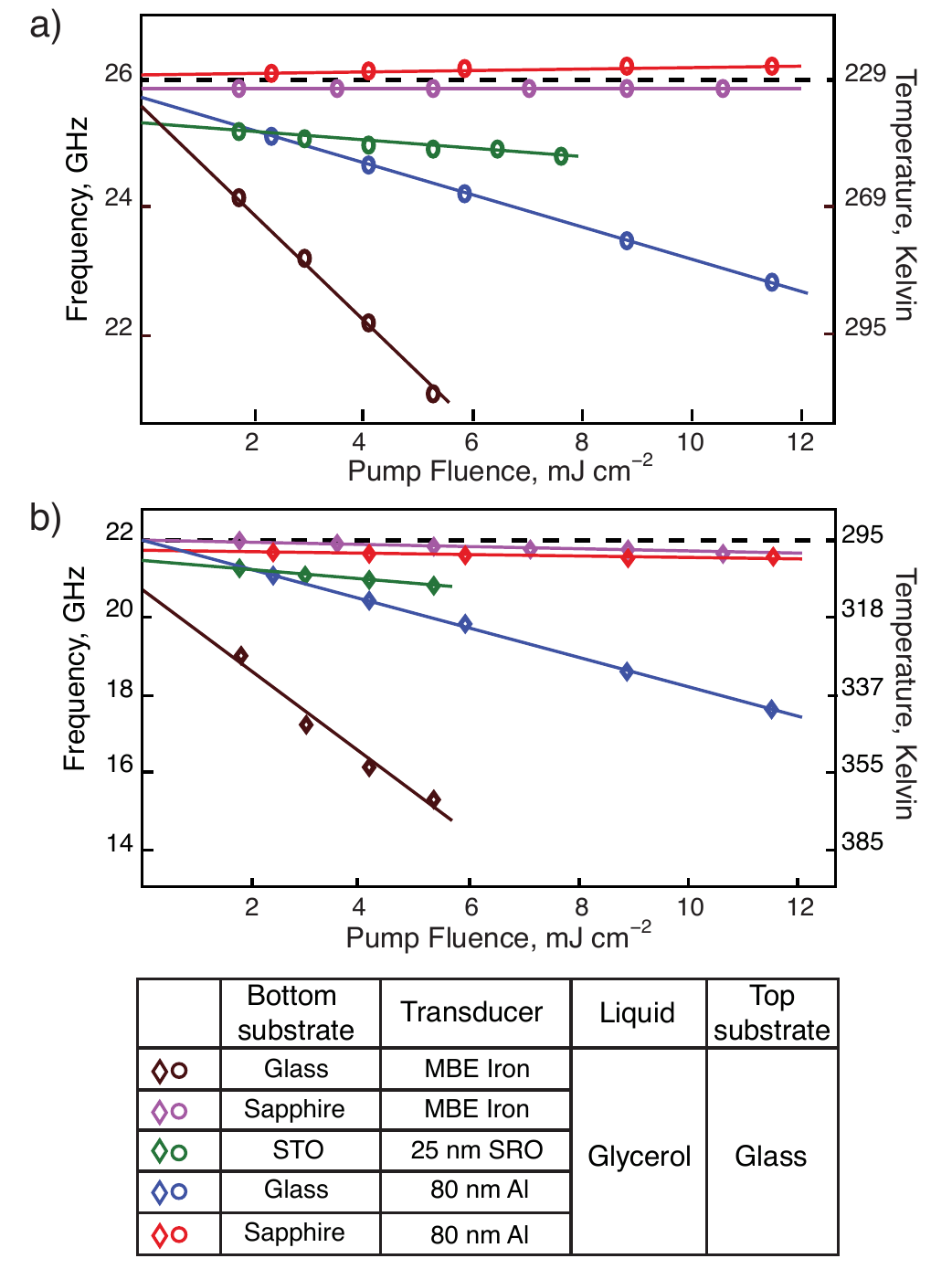}
\caption[Fig3]{(Color online) Brillouin scattering frequencies measured in glycerol at various laser fluences cause strongly different effects in cumulative sample heating for various sample structures. Different substrate materials holding different types of transducer films materials were studied. A transparent glass cover substrate and glycerol was used in all the sample configurations. Measurements on each sample assembly were made at two distinct temperatures of 229~K (a) and room temperature 295~K (b).The Brillouin scattering frequency were extracted from the recorded Brillouin scattering oscillations. The vertical temperature scales for both figures were deduced from the temperature calibration curve of Fig. \ref{fig2}(c).}
\label{fig3}
\end{figure}

Figure~\ref{fig3} shows the pump fluence dependence of the Brillouin scattering frequency for a variety of sample configurations at two different temperatures of the cryostat, 229~K and room temperature 295~K. Among the different sample configurations investigated, two samples have an 80~nm aluminum transducer film coated either on glass or sapphire substrates, two samples feature canted, molecular beam epitaxially (MBE) grown iron thin films, on either glass or sapphire substrates, and a sample with a 25~nm strontium ruthanate (SrRuO$_3$, SRO) thin film on an off-axis cut strontium titanate (SrTiO$_3$, STO) substrate. The MBE transducer films were used for shear wave generation as described in \cite{pezeril09, pezeril12, pezeril16, KHP+13}. The SRO sample was chosen because it attracted attention as a promising photoacoustic transducer material after a high laser-induced strain amplitude in a multilayer SRO structure was reported \cite{SRO, Maznev11}. In all configurations, the cover transparent substrate was glass and the liquid glycerol. The mean probe light intensity at each measurement was adjusted to account for differences in optical reflectivity among the different samples, such that the voltage on the detection photodiode was kept constant for all measurements. As shown in figure~\ref{fig3}, the Brillouin scattering frequency and the liquid temperature rise, deduced from Fig. \ref{fig2}(c), is strongly influenced by variation in pump fluence for a glass substrate, as opposed to the situation of sapphire or STO substrates. These discrepancies highlight the cumulative heating effect which manifests differently depending on the sample structures. Strong cumulative heating was always observed with a glass substrate holding a metallic transducer film deposited on it. The fluence-temperature slope, deduced from Fig. \ref{fig3}, was substantial in this sample configuration, about $\sim40$~K/10~mJ.cm$^{-2}$ for glass-aluminum and over $\sim100$~K/10~mJ.cm$^{-2}$ for glass-iron. The ratio of about 1:3 reflects the difference in optical absorbance of aluminum and iron at 790~nm of 13\% and 39\% respectively. In stark contrast to this was the cumulative heating observed in two sample configurations which have a transducer film on sapphire substrates. In this case, the pump fluence dependent heating was determined to be only 3~to~5~K/10~mJ.cm$^{-2}$. The explanation of this large difference in cumulative heating lies in the difference in thermal conductivity, $\kappa_{\textrm{\scriptsize glass}}\approx1.4\textrm{ W.m}^{-1}\textrm{.K}^{-1}$ and $\kappa_{\textrm{\scriptsize sapphire}}\approx45\textrm{ W.m}^{-1}\textrm{.K}^{-1}$ for glass and sapphire. Therefore, the large difference is simply due to the thermal conductivity of glass does not favor efficient thermal diffusion and dissipation, as compared to sapphire. Finally, SrTiO$_3$ is almost as good a thermal conductor as sapphire, $\kappa_{\textrm{\scriptsize STO}}\approx20\textrm{ W.m}^{-1}\textrm{.K}^{-1}$ when subjected to a heating of $\sim10$~K/10~mJ.cm$^{-2}$. The only difference compared to the previous samples is the large offset at zero pump power which was likely caused by strong absorption of the 395~nm probe light which was at least twice as much power for measurements on SrTiO$_3$/SrRuO$_3$ samples compared to the highly reflective metal films.

These differences clearly demonstrate the influence in the choice of sample structures for experiments where the liquid sample has to be kept undisturbed in order to proceed to non-invasive pump-probe measurements. In order to avoid strong cumulative thermal effects, materials with good thermal conductivity are required.

\subsection{Simulation of cumulative heating}

\begin{figure}[t!]
\centering
\includegraphics[width=9cm]{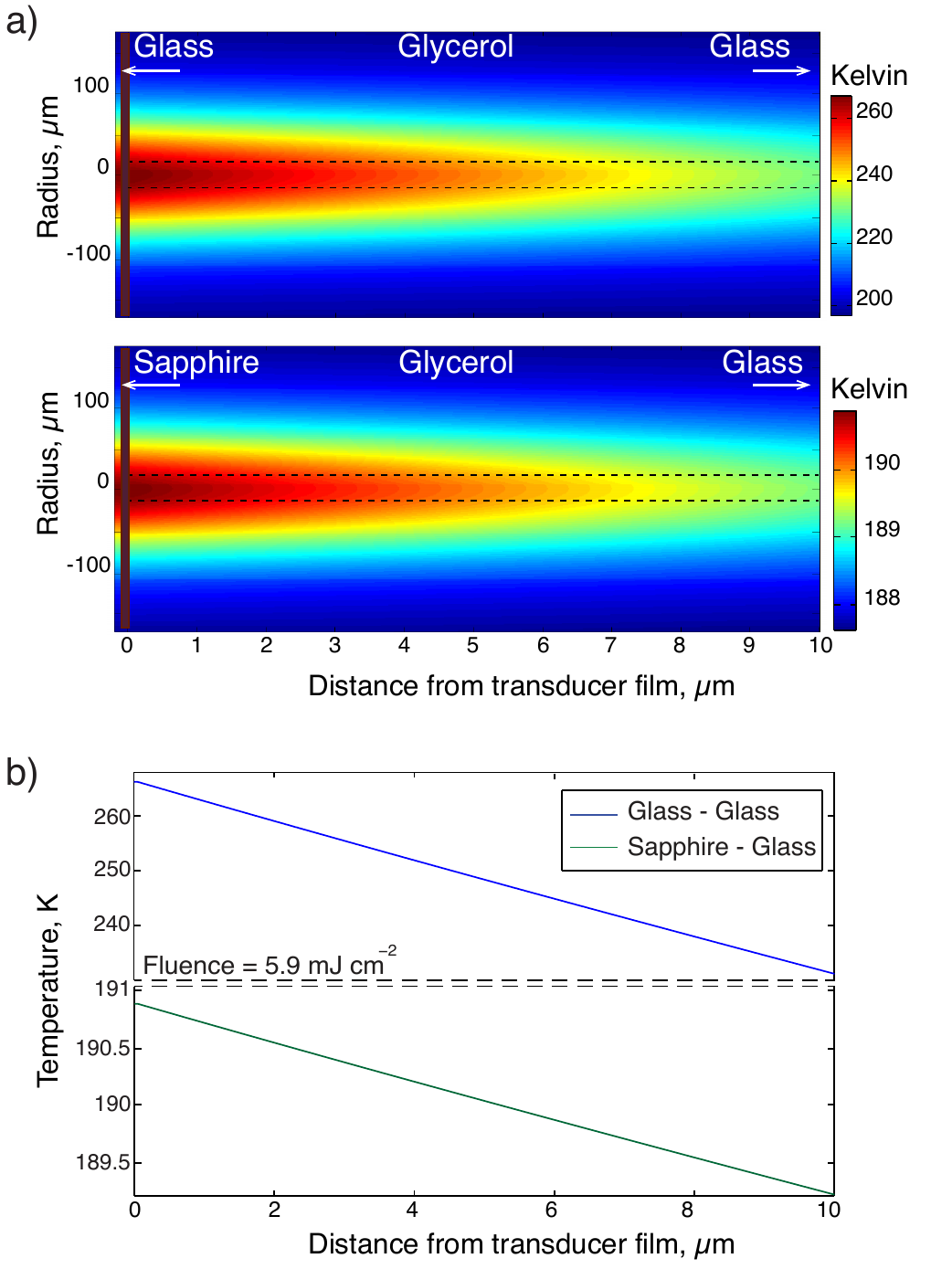}
\caption[Fig4]{(Color online) (a) 2D Results from simulations of cumulative heating in two different sample structures: a 50~nm iron transducer film, on either glass (Top) or sapphire (Bottom) substrate, which absorbs 39\% of a laser pump pulse of 5.9~mJ.cm$^{-2}$ fluence. The transparent cover substrate is glass. The resulting steady state heat distribution in a 10~microns glycerol film sandwiched between both substrates is calculated with the commercial software COMSOL$^\textrm{\scriptsize{\circledR}}$ Multiphysics. The glass transition temperature of glycerol T$_g$~=~186~K was chosen as the initial temperature in all the simulations. The dashed lines indicate the probe beam diameter. (b) Extracted temperature distribution plot from the 2D simulations displayed in (a). Note the linear temperature distribution and the pronounced different temperature scale and gradient in each plot.}
\label{fig4}
\end{figure}

In order to confirm our experimental observations, we have simulated the cumulative heating through modeling different sample structures and solving the time-independent heat equation in cylindrical geometry in COMSOL$^\textrm{\scriptsize{\circledR}}$ Multiphysics. The sample structure used in all simulations includes a substrate holding a 50~nm iron transducer film, glycerol, and a transparent cover substrate. An uniform absorption of 39\% of a 100~mW laser beam with a Gaussian beam profile and a FWHM spot size of 100~microns, which corresponds to a pump laser fluence of 5.9~mJ.cm$^{-2}$, throughout the depth of the transducer film was used as the heat input to the model system. Since the simulations were steady state and did not include the impulsive nature of heat deposition, the temperature profile obtained from these time-independent heat equation simulations gives an upper bound for the actual cumulative temperature rise.

Results from simulating several different substrate combinations are depicted in figure~\ref{fig4}(a) and (b). Acoustic waves at our detection wave vector propagate very far into the liquid, at least 5 to 10~microns, at low temperatures. Accordingly, the liquid thickness of glycerol in our simulations has been arbitrary chosen to be of 10~microns. The glass transition temperature of glycerol, T$_g$~=~186~K, was used as the initial temperature in all simulations. The top plot shows the 2D temperature distribution in a glass-glass configuration with a substantial temperature rise close to the transducer film of 84~K, the case with the highest temperature increase in our measurements. The corresponding temperature distribution within the probe spatial range, displayed in figure~\ref{fig4} (b), changes linearly by about 5~K/micron of propagation distance. In case of a sapphire-glass combination, see the bottom plot in figure~\ref{fig4} (b), our simulations confirm the fact that the temperature rise is more gentle, of about 15~K, in close agreement to our experimental observations. As opposed to a glass substrate, the linear temperature variation of a sapphire substrate is significantly reduced to 0.2~K/micron.

\begin{figure}[t!]
\centering
\includegraphics[width=9cm]{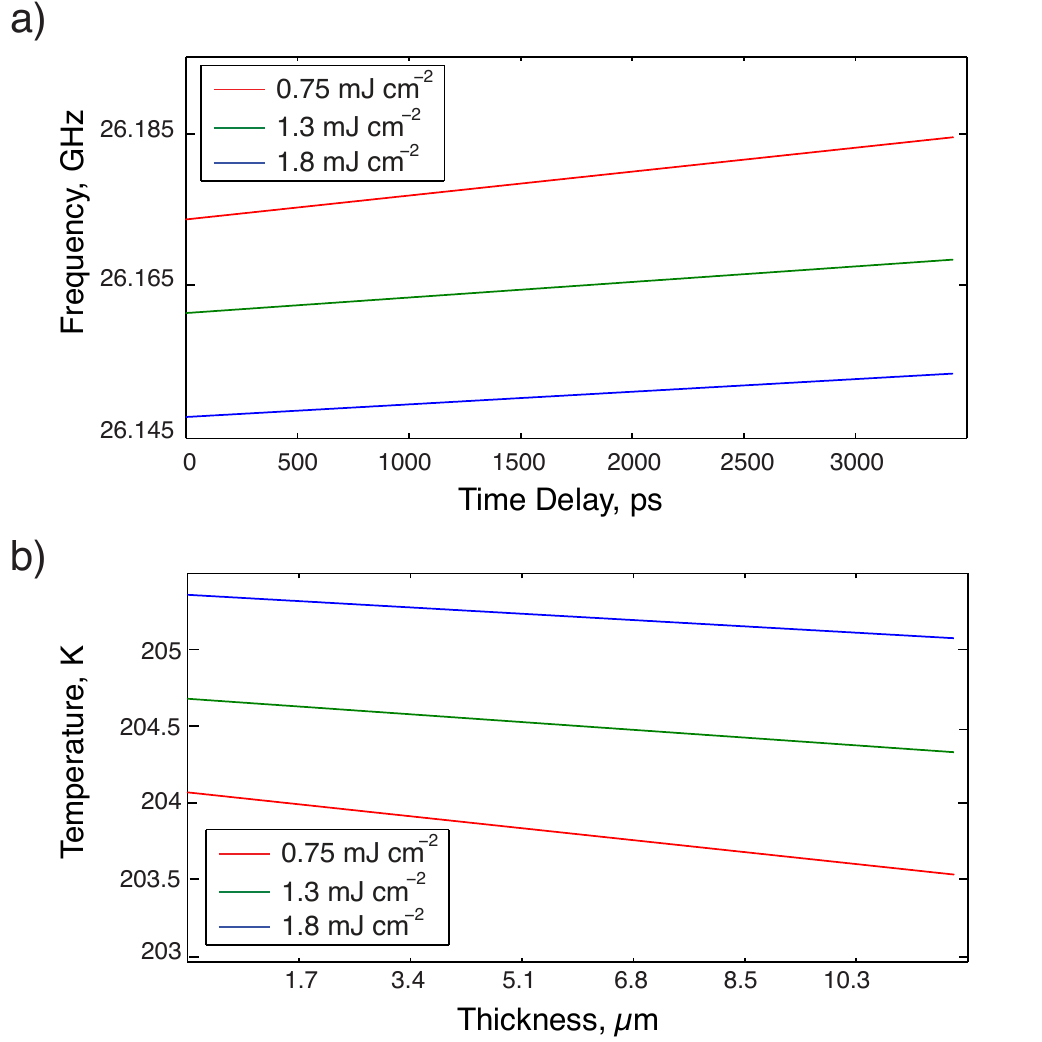}
\caption[Fig4b]{(Color online) (a) Time-resolved Brillouin frequency chirp fitted from our experimental results obtained from a 40~nm chromium transducer film deposited on a silicon substrate. The Brillouin frequency up-chirp matches the temperature distribution profile in glycerol at 200~K, for several laser fluences. (b) From the fitted Brillouin frequency in (a), we have obtained the temperature distribution through the liquid thickness, accordingly to the temperature calibration curve of Fig. \ref{fig2}(c).}
\label{fig4b}
\end{figure}

In order to experimentally reveal the in-depth temperature distribution in the liquid, we have examined further our TDBS data, searching for a frequency chirp in the measured Brillouin frequency. Accordingly to \cite{Klieber2015}, we have modified the functional form given in equation~(\ref{Brillouin}) by allowing the Brillouin frequency to vary in time, during the propagation of the acoustic pulse throughout the temperature gradient in the liquid. The modified Brillouin frequency functional form is then expressed as
                        \begin{equation}
                               \Delta\text{R}(t) \sim \: \sin ( 2\pi \nu(t) \, t + \phi ) \, \exp(-\Gamma t) \label{linear_chirp}
                        \end{equation}
where 
			\begin{equation}
			\nu(t) = \nu_0 + C \cdot t \label{eqnu}
			 \end{equation}			
represents a linearly chirped frequency, $\nu_0$ being the initial Brillouin frequency at time zero, i.e. close to the transducer film, and $C$ being the linear chirp coefficient. Even though a linear chirp does not reflect the true change in frequency, which has to model the temperature dependence of the Brillouin frequency, it provides a more stable algorithm and can be considered sufficient for the present analysis with little temperature variations across the liquid thickness. The extracted Brillouin frequency up-chirp from equation~(\ref{linear_chirp}) for a 40~nm chromium transducer film on a silicon substrate at 200~K and at several different laser fluences are depicted in figure~\ref{fig4b}(a) and (b). The results from the numerical fitting of our experimental data confirm that a temperature gradient is present in our liquid sample and that a good thermal conductor substrate such as silicon lead to weak cumulative laser heating effects, of only a couple of Kelvins. In addition, we notice that the Brillouin frequency decreases with an increase of pump fluence, which is in agreement with a stronger temperature rise at higher pump fluences.

\subsection{Multilayer structure for enhanced thermal insulation}

\begin{figure}[t!]
\centering
\includegraphics[width=9cm]{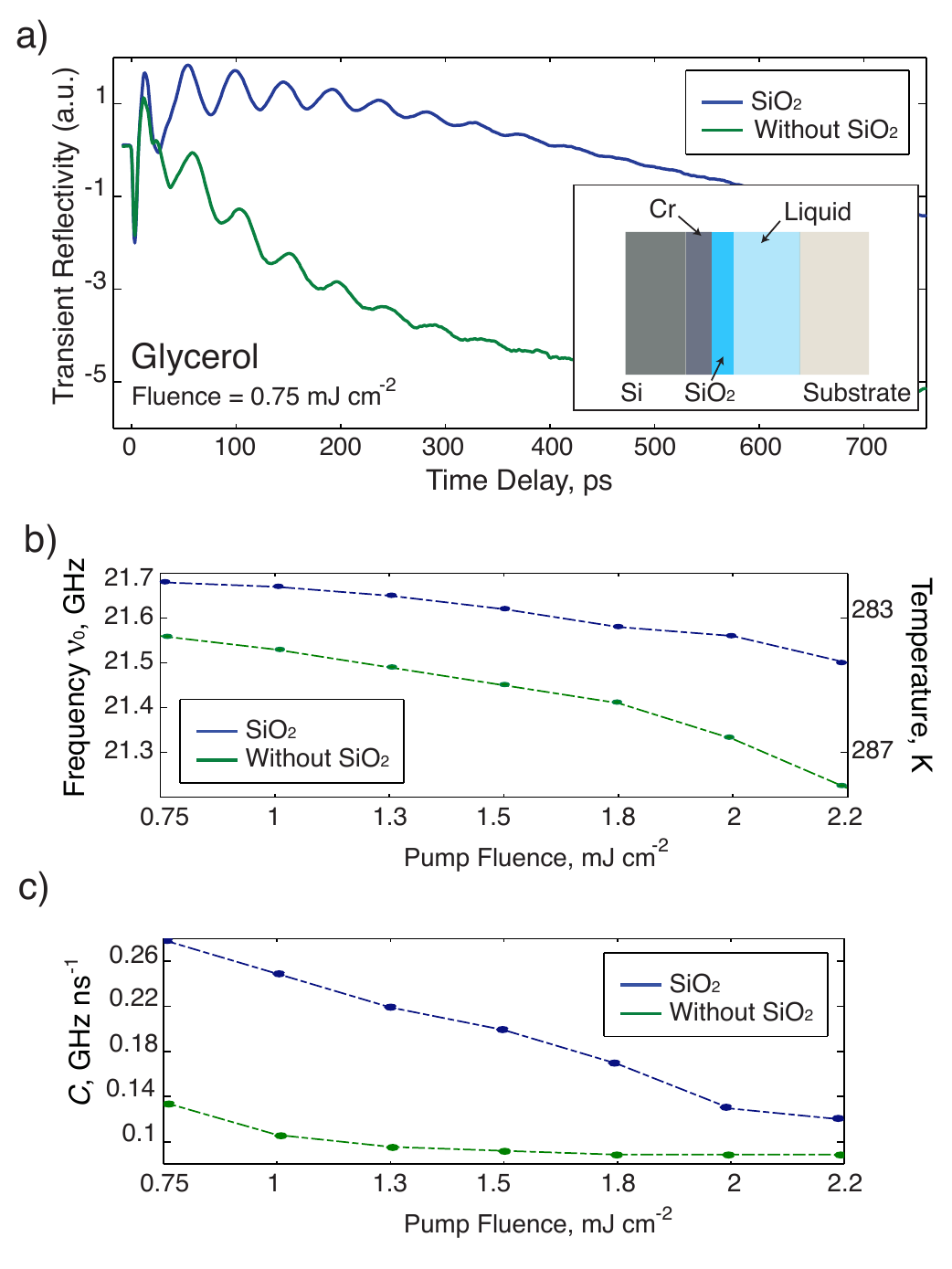}
\caption[SiO2layer]{(Color online) As sketched in the bottom inset of (a), a thin SiO$_2$ capping layer is introduced in order to thermally insulate the liquid from the laser heated Cr metallic transducer film. As before, a transparent glass cover substrate and glycerol was used in this specific sample configuration. (a) Results of the transient reflectivity signals measured in glycerol for both configurations, at a temperature of 280~K, as indicated by the Peltier temperature sensor. Evolution of the Brillouin frequency chirp coefficients $\nu_0$ (b) and $C$ (c) defined in Eq.(\ref{eqnu}) at several laser pump fluences, extracted from the transient reflectivity signals obtained in both configurations, with or without thermal insulator SiO$_2$ layer. The vertical temperature scale in (b) was deduced from the temperature calibration curve of Fig. \ref{fig2}(c). A noticeable discrepancy in (a) and (b) between both configurations reveals the benefit of the SiO$_2$ thermal insulator which prevents the liquid from cumulative laser heating. 
}
\label{fig5}
\end{figure}

In order to further decrease the influence of cumulative laser heating effects, which could be detrimental in sensitive experiments such as in studies of confined liquids \cite{Christenson1982, Heuberger2001, Perkin2012}, more complex sample configurations have to be explored. We have investigated a multilayer sample structure configuration where an additional thermal insulator layer is added as a thermal barrier between the hot metallic layer and the liquid. In this case, the sample has an additional silicon dioxide SiO$_2$ layer on top of the metallic film. First, the 40~nm chromium transducer film was deposited on a silicon substrate, and then a 19.7~nm SiO$_2$ layer was deposited on half of the metallic film for a straightforward comparison of the cumulative heating effect with or without silicon dioxide film, see inset of Fig.~\ref{fig5}(a). Measurements with this structure were made for a wide range of pump fluences for both sample configurations. Each recorded transient reflectivity data, see Fig.~\ref{fig5}(a) which corresponds to both situations, with or without SiO$_2$ buffer layer, was further processed in order to numerically extract the initial frequency $\nu_0$ and the chirp coefficient $C$ expressed in Eq.(\ref{eqnu}). The result of the fitting procedure for many different laser pump fluences is shown in Fig.~\ref{fig5}(b) and (c). We recall that the initial frequency $\nu_0$ reveals the change in liquid temperature whereas the chirp coefficient $C$ is related to the change in temperature gradient. Therefore, the frequency shift between both configurations, with or without SiO$_2$ layer, which appears in Fig.~\ref{fig5}(b) indicates the temperature rise in the liquid to be substantially minimized through the SiO$_2$ insulator layer. The temperature rise at the lowest fluence of 0.75~mJ.cm$^{-2}$ is only of 1.5~K with the SiO$_2$ insulator layer whereas more than twice as much at 3.6~K without the insulator layer. The efficiency of the insulator layer is clear, even at the highest pump fluence where the temperature rise is kept as little as 4.5~K, as compared to an uncapped chromium layer with 8.6~K temperature rise. The obvious difference in the thermoreflectance background signals in Fig.~\ref{fig5}(a) indicates that the transient temperature rise relaxes to thermal equilibrium quicker in case of an SiO$_2$ insulator layer, in agreement with our interpretation of the initial frequency chirp $\nu_0$ evolution with pump fluence in Fig.~\ref{fig5}(b). Note that since the optical constants mismatch between glycerol and the SiO$_2$ insulator layer is negligible, we can rule out any optical artifact in our interpretation of the data displayed in Fig.~\ref{fig5}(b) and (c).

The interpretation of the frequency chirp coefficient discrepancy between both configurations is not as straightforward. In Fig.~\ref{fig5}(c) we notice the chirp coefficient $C$ being higher for the SiO$_2$ insulator layer case, which means that the temperature gradient is more pronounced than without SiO$_2$ insulator layer. This behavior is inherent to the higher thermal effusivity $e$ of the SiO$_2$ insulator layer in comparison with glycerol. Since $e_{\text{SiO}_2} = 1.7\times10^{3}$ J.K$^{-1}$.m$^{-2}$.s$^{-1/2} > e_{\text{glycerol}} = 0.9\times10^{3}$ J.K$^{-1}$.m$^{-2}$.s$^{-1/2}$, i.e. $e_{\text{SiO}_2}/e_{\text{glycerol}}\approx 2$, the  SiO$_2$ layer increases the stationary heat flux from the optically heated chromium film into glycerol \cite{Incropera, Mandelis} that leads to the increase of the temperature gradient in glycerol. Therefore, the temperature rise in glycerol is localized at close vicinity of the chromium film. At the contrary, since the temperature gradient is weaker without insulating layer, the temperature rise is much less localized and homogeneously distributed in depth of the liquid sample. In Fig.~\ref{fig5}(c) we notice an up-chirp coefficient $C$ decrease with an increase in pump fluence, which means that the temperature gradient weakens at higher fluence for both sample configurations. We attribute this observation to the fact that at higher pump fluence the liquid temperature becomes more and more homogeneous, which causes the temperature gradient in the liquid to diminish with increasing pump fluence.

\section{Summary}

These experimental studies and simulations of different sample structures have shown the importance of the proper selection of sample substrates holding the metallic transducer films to avoid cumulative thermal heating effects. Such effects can be efficiently minimized by using a good thermally conducting substrate like sapphire or silicon. To further decrease the influence of cumulative thermal heating effects we have investigated a multilayer sample structure where a thermal insulating SiO$_2$ layer was added in order to shield the liquid from the laser heated metallic transducer film. We have experimentally demonstrated the benefit of this alternative sample structure which could be required in experimental situations where even slight temperature changes have to be avoided. 
\textcolor{black}{Ultimately, our results could shed light on the thermal properties of ultrathin confined liquid films \cite{Christenson1982, Heuberger2001, Perkin2012}. This is a fascinating experimental challenge for the understanding of nanoscale heat transport \cite{Cahill2003, Cahill2014,Volz2016}.}

\section*{ACKNOWLEDGEMENTS}

This work was partially supported by the Department of Energy under grant No.~DE-FG02-00ER15087, National Science Foundation under grants No.~CHE-0616939 and DMR-0414895, Agence Nationale de la Recherche under grant No.~ANR-12-BS09-0031-01. The authors acknowledge financial support from CNRS (Centre National de la Recherche Scientifique) under grant Projet International de Coop\'eration Scientifique. The authors would like to thank Lionel Guilmeau for technical support as well as Mathieu Edely for Chromium deposition.


\end{document}